\documentclass[twocolumn,groupedaddress,showpacs,nofootinbib]{revtex4-1}

\usepackage[colorlinks,pdfstartview=FitH]{hyperref}
\hypersetup{linkcolor=blue,citecolor=blue,filecolor=black,urlcolor=blue}

\usepackage{amsmath,amssymb,bm}
\usepackage{graphicx}
\usepackage{amsfonts}
\usepackage{color}

\usepackage{amsfonts}
\usepackage{bbm}
\usepackage{latexsym}

\begin{document}

\title{Dark leptophilic scalar with the updated muon $g-2$ anomaly}

\author{Lian-Bao Jia}
\email{jialb@mail.nankai.edu.cn}
\affiliation{
School of Science, Southwest University of Science and Technology, Mianyang 621010, China
}

\begin{abstract}

The muon $g-2$ anomaly was strengthened by recent experimental results at Fermilab, which may be signatures of new physics. A scenario of leptophilic scalar $\phi_L$ accounting for the muon $g-2$ anomaly is investigated in this paper. Though a light $\phi_L$ mainly decaying into standard model (SM) particles has been excluded by experiments, a dark leptophilic scalar $\phi_L$ predominantly decaying into invisible fermionic dark matter (DM) $\chi \bar{\chi}$ is still allowed. Considering the decay mode $\phi_L \to  \chi \bar{\chi}$ opened (here $m_{\phi_L} = 3 m_\chi$), the coupling preferred by the muon $g-2$ and the $\phi_L-\chi$ coupling are derived. The light/heavy $\phi_L$ (roughly 1 GeV as a benchmark value) can be tested by future experiments via DM/SM decay modes, and $\phi_L$'s contribution to the anomalous magnetic moment of tau lepton could be investigated at lepton colliders. The search of $\chi$ via $\chi-$electron scattering in DM direct detection is not sensitive due to a tiny $\phi_L-$electron coupling, especially for $m_\chi \gtrsim m_\mu$.

\end{abstract}

%\begin{document}

\maketitle
%\flushbottom
%\newpage

%%%%%%%%%%%%%%%%%%%%%%%%%%%%%%%%%%%%%%%%%%%%%%%
\section{Introduction}
\label{sec:Intro}
%%%%%%%%%%%%%%%%%%%%%%%%%%%%%%%%%%%%%%%%%%%%%%%

New physics beyond the standard model (BSM) may leave traces in low-energy precise measurements via possible deviations between the standard model (SM) and experimental observations. One long-standing discrepancy is the anomalous magnetic dipole moment of the muon, i.e. the quantity $a_\mu \equiv (g-2)_\mu/2$ measured at the Brookhaven National Laboratory \cite{Muong-2:2001kxu,Muong-2:2006rrc} is in tension with the SM predication with $\Delta a_\mu = a_\mu^\mathrm{EXP} - a_\mu^\mathrm{SM} = (27.9 \pm 7.6) \times 10^{-10}$ \cite{Aoyama:2020ynm}, corresponding to a 3.7$\sigma$ discrepancy. Recently, the first results from the Muon g-2 experiment at Fermilab confirms the tension \cite{Muong-2:2021ojo}, and the combined results are $\Delta a_\mu = (25.1 \pm 5.9) \times 10^{-10}$, up to 4.2$\sigma$ deviation.\footnote{The theoretical uncertainty is currently dominated by the hadronic vacuum polarization \cite{Crivellin:2020zul,Davier:2017zfy,Davier:2019can}, and the difference between the data-driven approach \cite{Aoyama:2020ynm} and recent lattice QCD result \cite{Borsanyi:2020mff} indicates that the anomaly of muon $g-2$ may also be moderate.} This anomaly further strengthens the evidence of new particles and new interactions BSM. Furthermore, the deviation of the electron anomalous magnetic moment between the measured value and the standard model prediction $\Delta a_e =  a_e^\mathrm{EXP} - a_e^\mathrm{SM}$ is $\Delta a_e^{[\mathrm{Cs}]} = (-87 \pm 36) \times 10^{-14}$  \cite{Parker:2018vye,Davoudiasl:2018fbb}, and $\Delta a_e^{[\mathrm{Rb}]} = (48 \pm 30) \times 10^{-14}$ \cite{Morel:2020dww}. The main uncertainty on $\Delta a_e$ at present is dominated by $a_e^\mathrm{EXP}$, and possible deviation of $\Delta a_e$ needs considered when evaluating new physics contributions to $\Delta a_\mu$.

Now, which kind of new physics (see e.g. Refs. \cite{Lindner:2016bgg,Dev:2017ftk,Liu:2020qgx,Bigaran:2020jil,Ghosh:2020fdc,Capdevilla:2020qel,Endo:2021zal,Yin:2021mls,Ge:2021cjz,Wang:2021fkn,Zhu:2021vlz,Wang:2021bcx,Cao:2021tuh,Keung:2021rps,Athron:2021iuf}) accounting for the muon anomaly and meanwhile being allowed by the relevant constraints becomes a crucial question. A simple extension for the anomaly is a light leptophilic scalar $\phi_L$ with enhanced coupling to leptons, and a promising search mode was suggested to be $e^+ e^- \to \tau^+ \tau^- \phi_L$ followed by $\phi_L \to l \bar l$ ($l = e, \mu$) \cite{Batell:2016ove} at high-luminosity electron-positron colliders. In practice, new results from the BABAR Collaboration \cite{BaBar:2020jma} have excluded such a leptophilic $\phi_L$ favored by the muon $g-2$ anomaly if $\phi_L$'s mass $m_{\phi_L}$ is below 4 GeV. In addition, the parameter spaces left with a larger mass of $m_{\phi_L} \sim$ 5$-$30 GeV and a larger coupling (dozens of times larger than the corresponding Higgs coupling) for the muon $g-2$ anomaly can be further judged at the future lepton collider designed as a Higgs factory \cite{Eung:2021bef}.

Another new physics BSM is the existence of dark matter (DM), while the microscopic properties of DM are largely unknown. If new interactions accounting for the muon anomaly are related to the dark sector, things will become interesting. In this paper, we focus on a dark leptophilic $\phi_L$ favored by the muon $g-2$ anomaly mediates the transitions between the dark sector and SM sector. For DM mass being lighter than $m_{\phi_L}$/2, the decay mode of $\phi_L \to$ a pair of DM particles is open. In this case, tensions from BABAR results \cite{BaBar:2020jma} can be removed when $\phi_L$'s decay products being predominantly invisible. Meanwhile, new sector particles should be compatible with the constraints from the big bang nucleosynthesis and the cosmic microwave background (CMB). For thermal freeze-out DM, the mass of DM particles should be $\gtrsim$ 10 MeV \cite{Ho:2012ug,Boehm:2013jpa,Jia:2016uxs,Berlin:2018sjs,Escudero:2018mvt}, and fermionic DM with $p-$wave annihilation evades constraints from CMB \cite{Slatyer:2015jla}. In addition, the $\phi_L -$electron coupling is suppressed compared with the $\phi_L -$muon coupling, and thus it can be allowed by the present direct detection experiments on DM-electron scattering \cite{XENON10:2011prx,Essig:2017kqs,DarkSide:2018ppu,XENON:2019gfn,SENSEI:2020dpa}. This dark leptophilic scalar $\phi_L$ will be explored in this paper.

%%%%%%%%%%%%%%%%%%%%%%%%%%%%%%%%%%%%%%%%%%%%%%%
\section{Framework of new interactions}
\label{sec:Model}
%%%%%%%%%%%%%%%%%%%%%%%%%%%%%%%%%%%%%%%%%%%%%%%

Here we consider a leptophilic scalar $\phi_L$ which is related to the dark sector as a solution to the muon $g-2$ anomaly. The effective interactions of $\phi_L$ with charged leptons are taken as
\begin{eqnarray}
\mathcal{L}_\mathrm{eff} \supset \frac{1}{2} (\partial_\mu \phi_L)^2  - \frac{1}{2} m_{\phi_L}^2 \phi_L^2  &-& \xi \sum_{\ell=e,\mu,\tau}  \frac{m_\ell}{v} \phi_L \bar{\ell} \ell  \; ,
\end{eqnarray}
where $\xi$ is the coupling strength relative to the SM Yukawa couplings $m_\ell / v$,
and $v =$ 246 GeV is the SM vacuum expectation value. The leptophilic scalar $\phi_L$ with mass-proportional interactions is UV-complete \cite{Batell:2016ove}, which can follow from a limit of the two-Higgs-doublet model with an additional singlet field. The one-loop contribution of $\phi_L$ to the $g - 2$ discrepancy of lepton $l$ ($l = \mu, e$) is \cite{Leveille:1977rc,Tucker-Smith:2010wdq,Chen:2015vqy}
\begin{eqnarray}
\Delta a_l =   \frac{ \xi^2}{8 \pi^2} \frac{m_l^2}{v^2} \int^1_0 dz \frac{(1+z)(1-z)^2}{(1-z)^2 +z (m_{\phi_L}/m_l)^2 }\; ,
\end{eqnarray}
which is related to the lepton's mass.

For a light $\phi_L$ with a mass less than a few GeV that can explain the muon $g-2$ anomaly, the $\phi_L$ with predominantly visible decay modes has been excluded by experiments, as addressed in the Introduction. The main SM decay modes of $\phi_L$ are $\phi_L \to e^+ e^-$, $\mu^+ \mu^-$ when $m_{\phi_L} > 2 m_\mu$ and $\tau^+ \tau^-$ when $m_{\phi_L} > 2 m_\tau$. The decay width of $\phi_L \to \ell \bar \ell$ is
\begin{eqnarray}
\Gamma_{\ell \bar \ell} =   \frac{ \xi^2}{8 \pi} \frac{m_\ell^2}{v^2} m_{\phi_L} (1 - \frac{4 m_\ell^2}{m_{\phi_L}^2})^{3/2} \; .
\end{eqnarray}
In addition, the decay width of $\phi_L \to \gamma \gamma$ is \cite{Djouadi:2005gi}
\begin{eqnarray}
\Gamma_{\gamma \gamma} =  \frac{ \xi^2 \alpha^2}{256 \pi^3} \frac{m_{\phi_L}^3}{v^2}  |\sum_{\ell=e,\mu,\tau} A_{1/2} (\tau_\ell)|^2 \; ,
\end{eqnarray}
with $\tau_\ell = m_{\phi_L}^2/4 m_\ell^2$, the form factor $A_{1/2} (\tau_\ell) = 2[\tau_\ell + (\tau_\ell - 1) f(\tau_\ell)]\tau_\ell^{-2}$, and the function $f(\tau_\ell)$ defined as
\begin{eqnarray}
 f(\tau_\ell) = \bigg \{ \begin{array}{cc}
 \arcsin^2\sqrt{\tau_\ell} ~~   \,  &  \tau_\ell \leq 1 \,, \\
  -\frac{1}{4}\big [  \log \frac{1+\sqrt{1- \tau_\ell^{-1} }}{1 - \sqrt{1- \tau_\ell^{-1} }} -i \pi   \big]^2 \,  &  \tau_\ell > 1 \,.   \nonumber
\end{array}
\end{eqnarray}
On the other hand, the scenario of $\phi_L$ predominantly decaying into invisible DM particles with DM mass $< m_{\phi_L}$/2 is still allowed by experiments, and this is of our concern. For fermionic DM $\chi$, the effective interaction can be written as
\begin{eqnarray}
\mathcal{L}_\mathrm{DM}^{int} = - g_D^{} \phi_L \bar{\chi} \chi \; .
\end{eqnarray}
The decay width of $\phi_L \to \chi \bar{\chi}$ is
\begin{eqnarray}
\Gamma_{ \chi \bar{\chi}} =   \frac{ g_D^2}{8 \pi} m_{\phi_L} (1 - \frac{4 m_\chi^2}{m_{\phi_L}^2})^{3/2} \; .
\end{eqnarray}

The annihilation cross section of $\chi \bar \chi$ into pairs of charged leptons and photons can be written as
\begin{eqnarray}
\sigma v_r =  \frac{s - 4m_D^2}{s - 2m_D^2} \frac{g_D^2}{2}  \frac{\sqrt{s}(\tilde{\Gamma}_{\ell \bar \ell} +\tilde{\Gamma}_{\gamma \gamma}) }{(s -m_{\phi_L}^2)^2 + m_{\phi_L}^2 \Gamma_{\phi_L}^2}  \; ,   \label{ann-cos}
\end{eqnarray}
where $v_r$ is the relative velocity between the two DM particles, and $s$ is the total invariant mass squared. $\Gamma_{\phi_L}$ is the decay width of $\phi_L$. $\tilde{\Gamma}_{\ell \bar \ell} $, $\tilde{\Gamma}_{\gamma \gamma}$ are the rate of the virtual $\phi_L$ transiting into the SM particles, with the mass $m_{\phi_L}$ replaced by $\sqrt{s}$ in the calculations. In the non-relativistic limit, $s = 4 m_\chi^2 + m_\chi^2 v_r^2 + \mathcal{O} (v_r^4)$ is obtained. The annihilation of $\chi \bar \chi$ is $p-$wave dominant during the freeze-out period, and the relic density of $\chi \bar \chi$ can be derived by the usual calculations \cite{Gondolo:1990dk,Kolb:1990vq,Griest:1990kh}. In the following, a numerical analysis of $\phi_L$ mediated fermionic DM will be proceeded with the updated results of the muon $g-2$ anomaly.

%%%%%%%%%%%%%%%%%%%%%%%%%%%%%%%%%%%%%%%%%%%%%%%
\section{Analysis and test}
\label{sec:GCE}
%%%%%%%%%%%%%%%%%%%%%%%%%%%%%%%%%%%%%%%%%%%%%%%

\begin{figure}[htbp!]
\includegraphics[width=0.45\textwidth]{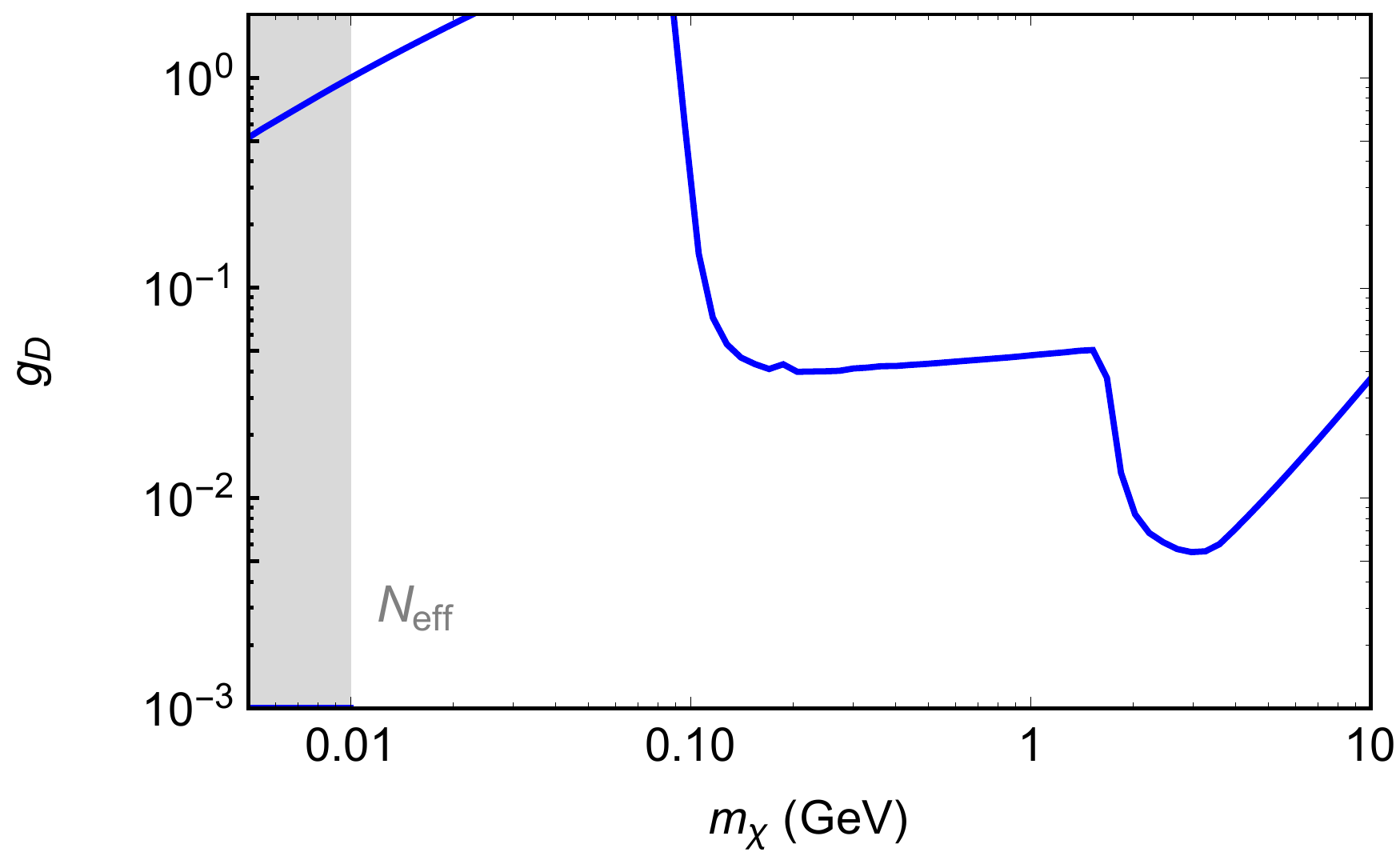} \vspace*{-1ex}
\caption{The coupling parameter $g_D^{}$ as a function of $m_\chi$ with $m_{\phi_L} = 3 m_\chi$ adopted. The value of $g_D^{}$ is derived by demanding that the relic density of $\chi \bar \chi$ is equal to today's DM observation value $\Omega_D h^2 =0.120$ \cite{Planck:2018vyg}, with two modes of $\chi \bar \chi$ annihilating into $\ell \bar \ell$ and $\gamma \gamma$ considered. The shaded region is possible bounds on DM masses from the effective number of relativistic neutrinos $N_{\mathrm{eff}}$ \cite{Berlin:2018sjs,Escudero:2018mvt}.}
\label{dcoup}
\end{figure}

\begin{figure}[htbp!]
\includegraphics[width=0.45\textwidth]{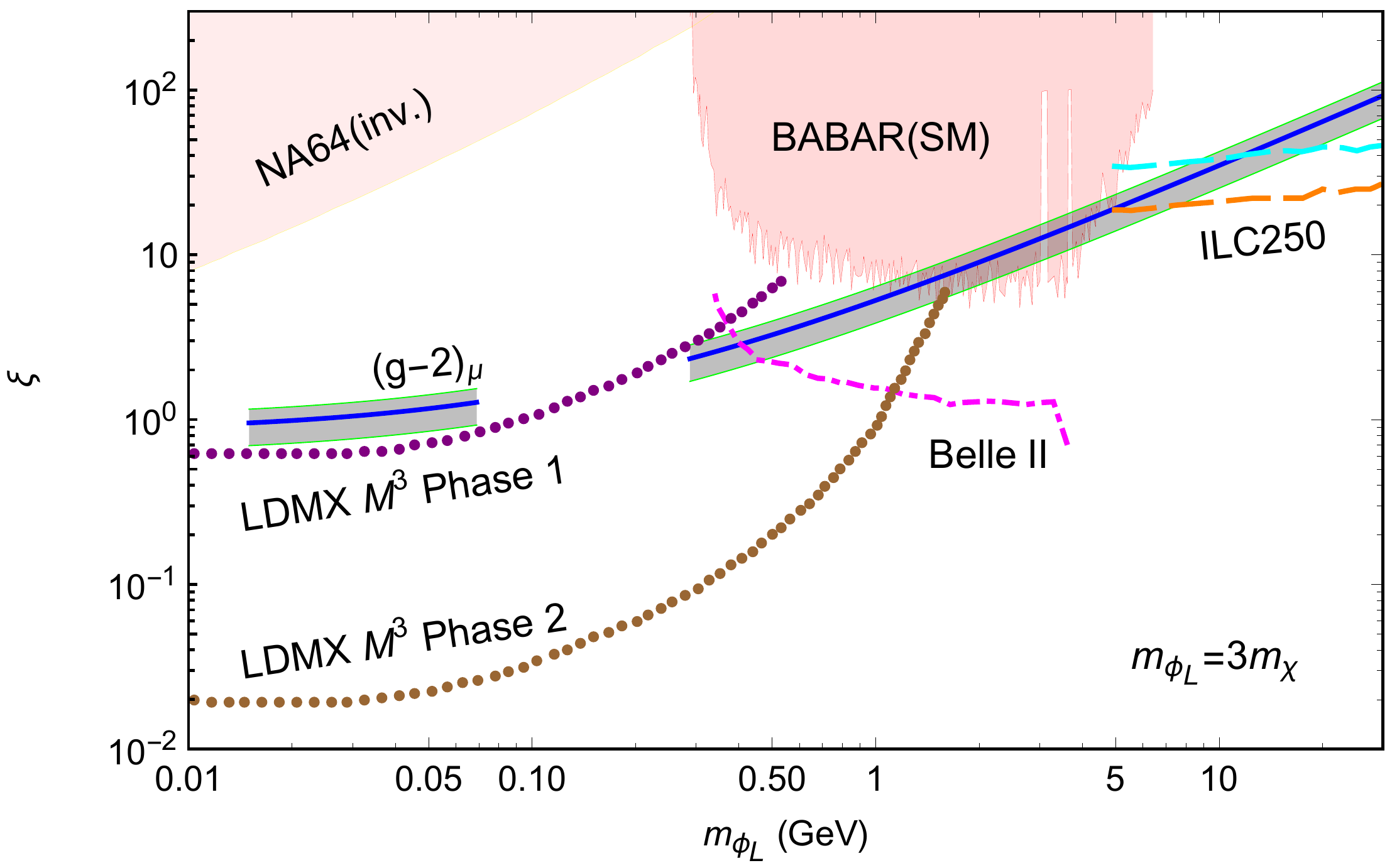} \vspace*{-1ex}
\caption{The coupling parameter $\xi$ as a function of $m_{\phi_L}$ preferred by the muon $g-2$ anomaly (two bands), together with existing constraints (shaded regions) of NA64 \cite{NA64:2021xzo} and revised BABAR limits \cite{BaBar:2020jma}, and projected search sensitivities. The projections of LDMX \cite{Berlin:2018bsc} and Belle II (revised results) \cite{Batell:2016ove} are shown by the dotted and the dot-dashed curves, respectively. Limits from ILC with 2$ab^{-1}$ data at 250 GeV \cite{Eung:2021bef} are shown by the dashed curves, with the lower (upper) one corresponding to the exclusion (discovery) limit.  }
\label{phi-xi}
\end{figure}

The parameter $\xi$ of the $\phi_L -\ell$ coupling is set by the muon $g-2$ results. Assuming that the main component of DM is $\chi \bar \chi$, the $\phi_L -\chi$ coupling parameter $g_D^{}$ can be obtained. Taking the relic density of DM today $\Omega_D h^2 =0.120 \pm 0.001$ \cite{Planck:2018vyg}, the results of the coupling parameter $g_D^{}$ is derived with a benchmark choice of $m_{\phi_L} = 3 m_\chi$, as shown in Fig. \ref{dcoup}. In the perturbative limit, the region with the coupling parameter $g_D^{} <2$ is considered here, and we have $m_\chi \lesssim 23$ MeV, or $m_\chi \gtrsim 96$ MeV. Hence, the mass range of $\phi_L$ of concern is $m_{\phi_L} \lesssim 69$ MeV, or $m_{\phi_L} \gtrsim 288$ MeV for $m_{\phi_L} = 3 m_\chi$. The the parameter space of $\xi$ as a function of $m_{\phi_L}$ preferred by the $(g-2)_\mu \pm 2\sigma$ is depicted as the two bands in Fig. \ref{phi-xi}.

\begin{figure}[htbp!]
\includegraphics[width=0.45\textwidth]{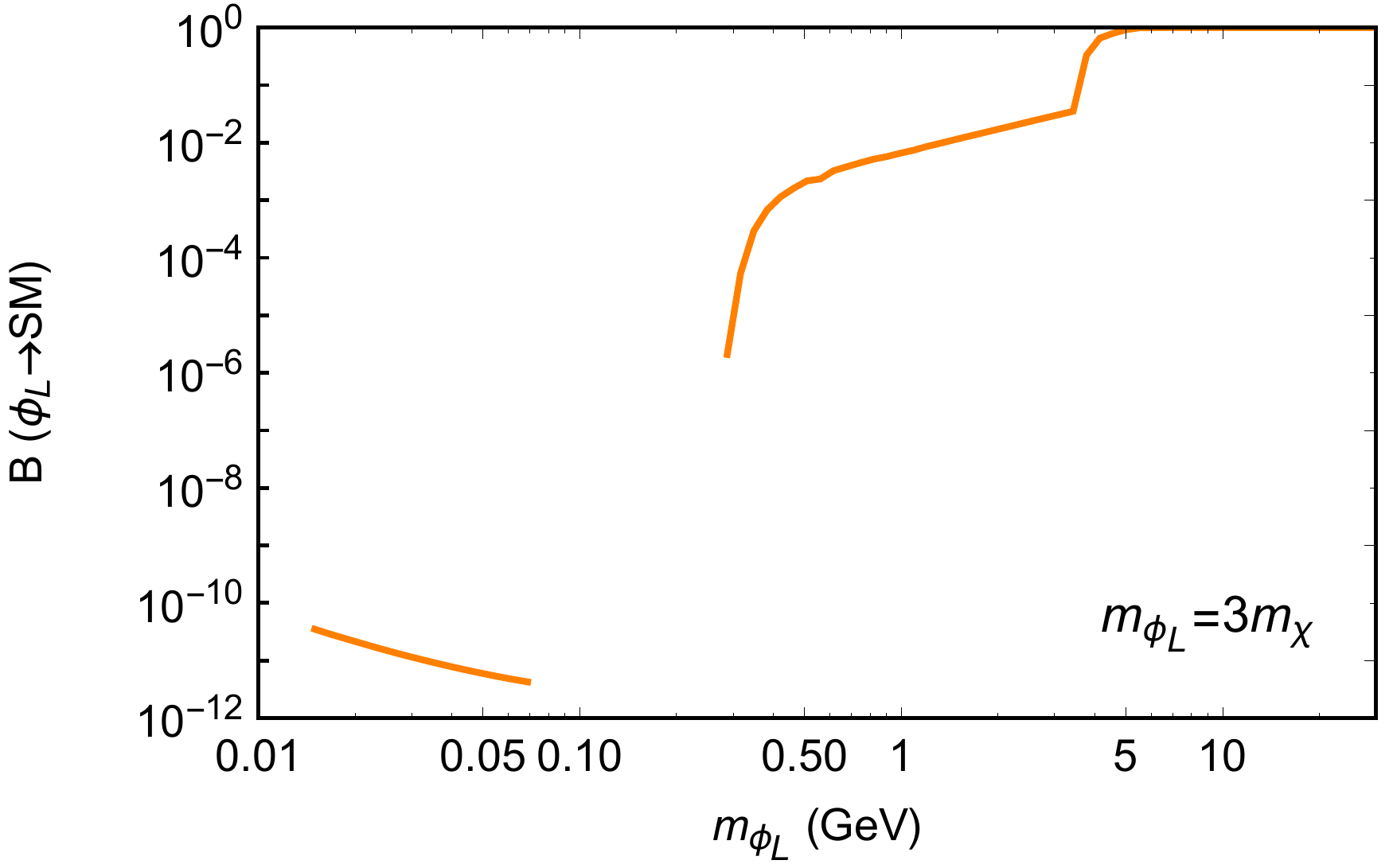} \vspace*{-1ex}
\caption{The branching ratio $\mathcal{B}$($\phi_L \to$ SM) of $\phi_L$ decaying into SM particles with $m_{\phi_L} = 3 m_\chi$ adopted.}
\label{phi-sm}
\end{figure}

The way of searching for the $\phi_L$ suggested by the muon $g-2$ anomaly depends on the main decay modes of $\phi_L$. The two decay modes of $\phi_L \to  \chi \bar{\chi}$ and $\phi_L \to$ SM particles varies with $\phi_L$'s mass, and the SM branching ratio $\mathcal{B}$($\phi_L \to$ SM) is shown in Fig. \ref{phi-sm}, with $m_{\phi_L} = 3 m_\chi$. $\phi_L$ can be probed in different mass ranges.

a: In the case of $m_{\phi_L} \lesssim$ 2 GeV, the DM decay mode $\phi_L \to  \chi \bar{\chi}$ is dominant, and this invisible $\phi_L$ can be searched at beam dump experiments via the missing energy. One such experiment is NA64 with 100 GeV electron scattering off nuclei, and the corresponding constraint \cite{NA64:2021xzo} is shown in Fig. \ref{phi-xi}, which is above the region favored by the the muon $g-2$ anomaly. The projections of LDMX Phase 1 and Phase 2 \cite{Berlin:2018bsc} with a muon missing momentum (M3) search is also shown in Fig. \ref{phi-xi}, which is sensitive to the range of $m_{\phi_L} \lesssim$ 1 GeV.

b: In the range of 0.5 GeV $\lesssim m_{\phi_L} \lesssim$ 5 GeV, the branching ratio $\mathcal{B}$($\phi_L \to$ SM) is $\gtrsim 2 \times 10^{-3}$. The $\phi_L$ without $\phi_L \to  \chi \bar{\chi}$ mode was suggested to be searched via the process $e^+ e^- \to \tau^+ \tau^- \phi_L$ followed by $\phi_L \to l \bar l$ ($l = e, \mu$) at high-luminosity electron-positron colliders \cite{Batell:2016ove}. Here the decay products $\phi_L \to l \bar l$ are suppressed by a factor of the branching ratio $\mathcal{B}$($\phi_L \to$ SM), and the revised BABAR limits \cite{BaBar:2020jma} and the projected sensitivity of Belle II \cite{Batell:2016ove} are presented in Fig. \ref{phi-xi}. For 0.5 GeV $\lesssim m_{\phi_L} \lesssim$ 5 GeV, some parameter spaces have been excluded by BABAR, and the left parameter spaces can be tested by Belle II.

c: For $m_{\phi_L} \gtrsim$ 5 GeV, the branching ratio $\mathcal{B}$($\phi_L \to$ SM) is close to 1 despite the mode $\phi_L \to  \chi \bar{\chi}$ allowed. The $\phi_L$ in this range can be investigated at the future lepton collider as a Higgs factory \cite{Eung:2021bef}. \\
Thus, the $\phi_L$ suggested by the muon $g-2$ anomaly can be tested by the future experiments.

%\begin{figure}[htbp!]
%\includegraphics[width=0.45\textwidth]{electron-g-2.pdf} \vspace*{-1ex}
%\caption{$\phi_L$'s contribution to $\Delta a_e$ (the two solid curves). The dashed curve is a typical value of $\Delta a_e = 5 \times 10^{-13}$ as an upper bound. }
%\label{e-g-2}
%\end{figure}

In addition, the supernova cooling may set constraints on new light bosons with masses in MeV \cite{Bollig:2020xdr,Croon:2020lrf}. For the new particle $\phi_L$ considered in this paper, $\phi_L$ couples to DM, and MeV $\phi_L$ mainly decays into $\chi \bar{\chi}$. Due to a considerable $\phi_L$-$\chi$ coupling (Fig. \ref{dcoup}), the inverse decay process of $\chi \bar{\chi} \to \phi_L$ dominates the DM absorption, and the dark sector luminosity is much smaller than the neutrino luminosity (see e.g., Ref. \cite{Sung:2021swd}). Thus, the coupling parameters of concern are not limited by the energy loss from supernova.

Besides the muon $g-2$ anomaly induced by $\phi_L$, here let's give an estimate of $\phi_L$'s contribution to the electron $g-2$ ($\Delta a_e$) first, then turn to the tau lepton anomalous magnetic moment ($\Delta a_\tau$). Considering the uncertainty of the experimental value $a_e^\mathrm{EXP}$ mentioned above, here we take a typical value of $|\Delta a_e| \lesssim 5 \times 10^{-13}$ as a bound of $\phi_L$'s contribution. Substituting the coupling parameter $\xi$ set by the muon $g-2$ anomaly, the result of $\phi_L$'s contribution to $\Delta a_e$ can be obtained, i.e., $\phi_L$'s contribution to $\Delta a_e$ is at least $\sim$ 3 orders of magnitude below the current precision (the typical bound). For tau lepton, constraints on $\Delta a_\tau$
is mild, with  $-0.007 < \Delta a_\tau^{\mathrm{BSM}} <$ 0.005 \cite{Gonzalez-Sprinberg:2000lzf}. Considering $\phi_L$'s contribution to the tau $g-2$, the anomalous magnetic moment $\Delta a_\tau^{\mathrm{BSM}}$ is up to $\sim 10^{-6} - 10^{-4}$ when the parameters are fixed to explain the muon $g-2$ discrepancy. Hence, $\phi_L$'s contribution to $\Delta a_\tau^{\mathrm{BSM}}$ could be within the reach of an upgraded B-factory with polarized electrons \cite{Crivellin:2021spu}.

\begin{figure}[htbp!]
\includegraphics[width=0.45\textwidth]{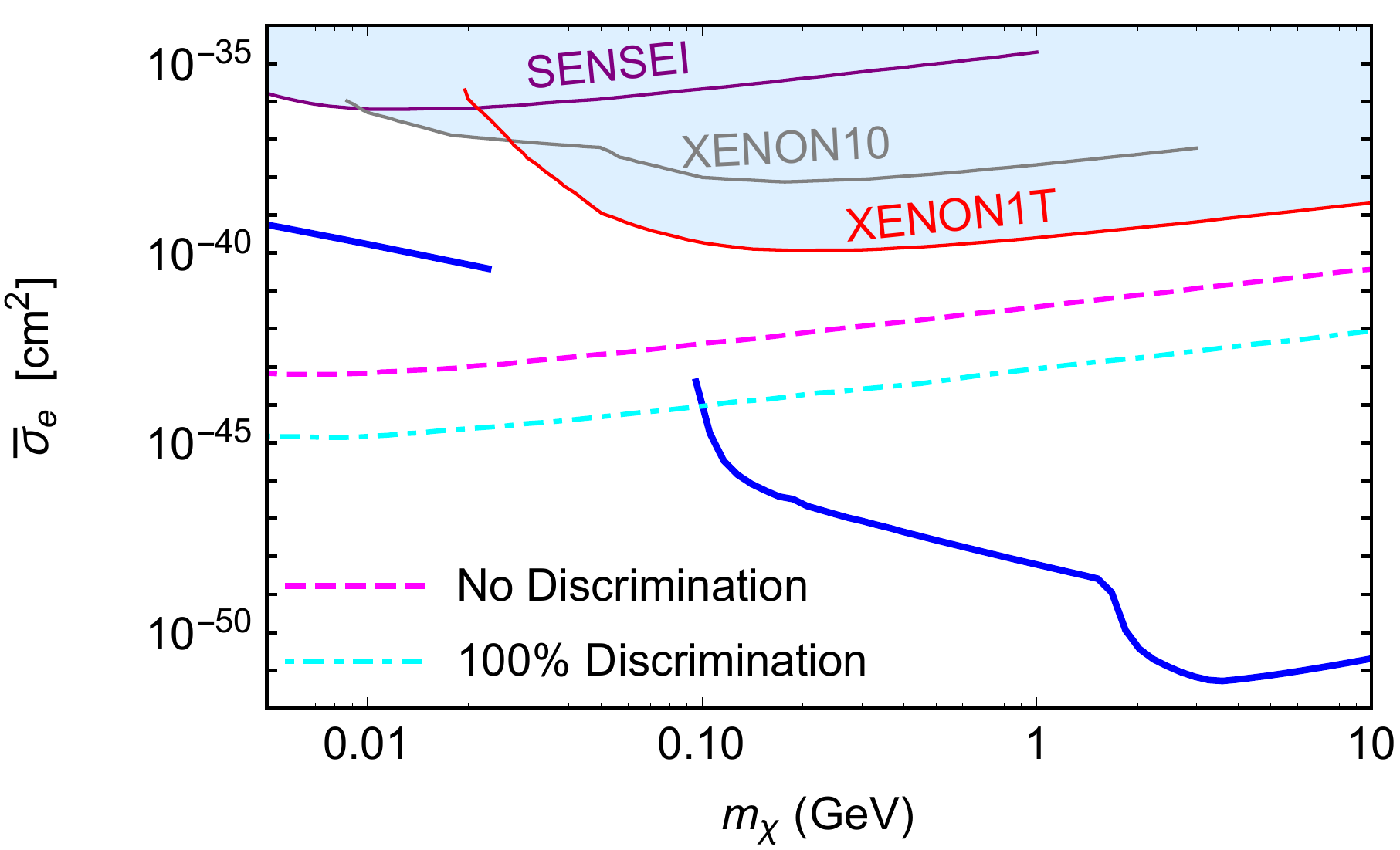} \vspace*{-1ex}
\caption{The cross section $\bar{\sigma}_{\rm e }^{}$ as a function of $m_\chi$ with $m_{\phi_L} = 3 m_\chi$. The two solid curves are the cross sections set by the muon $g-2$ anomaly. The shaded regions are the upper limits from SENSEI \cite{SENSEI:2020dpa}, XENON10 \cite{Essig:2017kqs} and XENON1T \cite{XENON:2019gfn}. The dashed and dot-dashed curves are the neutrino backgrounds for an exposure of $10^5$ kg-yr \cite{Wyenberg:2018eyv} with no discrimination and 100\% discrimination between electronic and nuclear scattering events, respectively.}\label{dm-electron}
\end{figure}

Now we give a brief discussion about the direct detection of DM particle $\chi$. Due to a tiny coupling between electron and $\phi_L$, the direct detection of $\chi$ via $\chi-$electron scattering is not sensitive, especially for a larger DM mass of $m_\chi \gtrsim m_\mu$. The $\chi-$electron scattering cross section is
\begin{eqnarray}
\bar{\sigma}_{\rm e }^{} \simeq   \frac{ g_D^2 \xi ^2 m_e^2}{\pi v^2 m_{\phi_L}^4} {m_\chi^2 m_e^2 \over (m_\chi + m_e)^2}  \; ,     \label{dm-e-scatt}
\end{eqnarray}
with the form factor $F_\mathrm{DM}(q) = 1$~\cite{Essig:2011nj}. Substituting the couplings set by the muon $g-2$ results into Eq. (\ref{dm-e-scatt}), the result of $\bar{\sigma}_{\rm e }^{}$ is shown in Fig. \ref{dm-electron}, which is below the present constraints from SENSEI \cite{SENSEI:2020dpa}, XENON10 \cite{Essig:2017kqs} and XENON1T \cite{XENON:2019gfn}. The discovery reach of DM direct detection experiments is limited by the neutrino background. For the neutrino background in $\chi-$electron scattering, the discovery limits vary with the exposure, electronic/nuclear recoil discrimination (traditional methods fail for nuclear and electronic recoil discrimination at low energy). The neutrino backgrounds for an exposure of $10^5$ kg-yr \cite{Wyenberg:2018eyv} with no discrimination and 100\% discrimination between electronic and nuclear scattering events are presented in Fig. \ref{dm-electron}. For $m_\chi \lesssim 23$ MeV, the $\bar{\sigma}_{\rm e }^{}$ is above the neutrino background. For $m_\chi \gtrsim 96$ MeV, the $\bar{\sigma}_{\rm e }^{}$ is below the neutrino background in detections.

%%%%%%%%%%%%%%%%%%%%%%%%%%%%%%%%%%%%%%%%%%%%%%%
\section{Conclusion}
\label{sec:Con}
%%%%%%%%%%%%%%%%%%%%%%%%%%%%%%%%%%%%%%%%%%%%%%%

We have investigated a scenario of dark leptophilic scalar $\phi_L$ accounting for the muon $g-2$ anomaly confirmed by the Fermilab Muon g-2 experiment. Though a light leptophilic scalar (below 4 GeV) mainly decaying into SM particles has been excluded by experiments, a dark leptophilic scalar $\phi_L$ predominantly decaying into invisible DM particles $\chi \bar{\chi}$ is still allowed. Considering the case of the decay mode $\phi_L \to  \chi \bar{\chi}$ opened (here $m_{\phi_L} = 3 m_\chi$), the coupling parameter $\xi$ preferred by the muon $g-2$ and the $\phi_L-\chi$ coupling are derived. For $m_{\phi_L} \lesssim$ 2 GeV, the DM decay $\phi_L \to  \chi \bar{\chi}$ is dominant. In the range of 0.5 GeV $\lesssim m_{\phi_L} \lesssim$ 5 GeV, the branching ratio $\mathcal{B}$($\phi_L \to$ SM) is $\gtrsim 2 \times 10^{-3}$. For $m_{\phi_L} \gtrsim$ 5 GeV, the branching ratio $\mathcal{B}$($\phi_L \to$ SM) is close to 1. The $\phi_L$ suggested by the muon $g-2$ anomaly can be tested by future experiments such as LDMX \cite{Berlin:2018bsc}, Belle II \cite{Batell:2016ove} and ILC \cite{Eung:2021bef}. Moreover, $\phi_L$'s contribution to the $\Delta a_\tau^{\mathrm{BSM}}$ of tau lepton is up to $\sim 10^{-6} - 10^{-4}$, which could be tested by an upgraded B-factory with a polarized electron beam. Due to the $\phi_L-$electron coupling being tiny, the search of $\chi$ via $\chi-$electron scattering in DM direct detection is not sensitive, especially for $m_\chi \gtrsim m_\mu$. We look forward to the further investigation of the muon $g-2$ both in theory and experiment.

%%%%%%%%%%%%%%%%%%%%%%%%
\acknowledgments
%%%%%%%%%%%%%%%%%%%%%%%%

The work of L.-B. Jia acknowledges support of the Longshan academic talent research supporting program of SWUST under Contract No. 18LZX415.

%%%%%%%%%%%%%%%%%%%%%%%%%%%%%%%%%%%%%

\end{document}